\documentclass{aa}
\usepackage{appendix}
\usepackage{latexsym}
\usepackage{graphicx}

\begin{document}

\title{The role of low-mass star clusters in massive star formation. The Orion Case.}
\author{V. M. Rivilla\inst{1} \and J. Mart\'in-Pintado\inst{1} \and I. Jim\'enez-Serra\inst{2} \and A. Rodr\'iguez-Franco\inst{1}}
\institute{Centro de Astrobiolog\'ia (CSIC-INTA), Ctra. de Torrej\'on-Ajalvir, km. 4, E-28850 Torrej\'on de Ardoz, Madrid, Spain\\
             \email{rivilla@cab.inta-csic.es}\\
      \and
           Harvard-Smithsonian Center for Astrophysics, 60 Garden St., 02138, Cambridge, MA, USA\\  }

 %  \date{Received _ ; accepted _ }
\date{Received 16/06/2011; accepted 28/01/2013}

  \abstract
  {Different theories have been proposed to explain the formation of massive stars: two are based on accretion processes (monolithic core accretion and competitive accretion), and another on coalescence of low- and intermediate-mass stars. To distinguish between these theories, it is crucial to establish the distribution, the extinction, and the density of low-mass stars in massive star-forming regions.}
   {Our aim is to analyze deep X-ray observations of the Orion massive star-forming region using the Chandra Orion Ultradeep Project (COUP) catalog, to reveal the distribution of the population and clustering of PMS low-mass stars, and to study their possible role in massive star formation.}
  {We studied the distribution of PMS low-mass stars with X-ray emission in Orion as a function of extinction with two different methods: a spatial gridding and a close-neighbors method, with cells of $\sim$ 0.03 $\times$ 0.03 pc$^2$, the typical size of protostellar cores. We derived density maps of the stellar population and calculated cluster stellar densities.}
 {We found that PMS low-mass stars cluster toward the three massive star-forming regions: the Trapezium Cluster (TC), the Orion Hot Core (OHC), and the OMC1-S region. We derived PMS low-mass stellar densities of 10$^5$ stars pc$^{-3}$ in the TC and OMC1-S, and of 10$^6$ stars pc$^{-3}$ in the OHC. The close association between the low-mass star clusters with massive star cradles supports the role of these clusters in the formation of massive stars. The X-ray observations show for the first time in the TC that low-mass stars with intermediate extinction are clustered toward the position of the most massive star $\theta^1$ Ori C, which is surrounded by a ring of non-extincted PMS low-mass stars. This 'envelope-core' structure is also supported by infrared and optical observations. Our analysis suggests that at least two basic ingredients are needed in massive star formation: the presence of dense gas and a cluster of low-mass stars. The scenario that better explains our findings assumes high fragmentation in the parental core, accretion at subcore scales that forms a low-mass stellar cluster, and subsequent competitive accretion. Finally, although coalescence does not seem a common mechanism for building up massive stars, we show that a single stellar merger may have occurred in the evolution of the OHC cluster, favored by the presence of disks, binaries, and gas accretion.} 
{}   % conclusions heading (optional), leave it empty if necessary 

\authorrunning{V. M. Rivilla et al.} 
\maketitle

\section{Introduction}

Massive stars (M$>$8M$_{\odot}$) are the main source of mechanical energy injection into the interstellar medium of galaxies. Although they are important for the evolution of galaxies, their formation process is not clearly understood. 
This is due to the large distances to the massive star-forming regions, their short evolution time-scales, the complexity of their birth regions and the high extinction.
Several theories have been proposed to explain the formation of
the most massive stars: i) monolithic gravitational collapse and core accretion (Yorke $\&$ Sonnhalter, \cite{yorke02}; McKee $\&$ Tan, \cite{mckee03}; Krumholz et al., \cite{krumholz09}); ii) coalescence of low-mass stars in high stellar density clusters (Bally $\&$ Zinnecker, \cite{bally05}; Bonnell et al., \cite{bonnell}; Stahler et al., \cite{stahler00}); and iii) competitive accretion of low- and intermediate-mass stars in the cluster potential well (Bonnell $\&$ Bate, \cite{bonnell06}). 
While the classic theory of monolithic core accretion predicts the formation of
mainly massive companions in small multiple systems, coalescence and competitive accretion models require a dense population of low-mass stars that form a cluster. 
 Therefore, to distinguish between the different scenarios of massive star formation, it is crucial to establish the distribution and density of low-mass stars in massive star cradles.

Measuring the population of low-mass stars where the early phases of massive star formation 
take place has been extremely difficult because they lie behind high extinction.
X-ray observations are particularly useful for studying high-mass star-forming regions because they can penetrate into the cloud despite the high extinctions found in these regions ($>$15 mag). In addition, they suffer less from foreground- and background galactic contamination than optical or infrared (IR) studies, and allow a census of pre-main sequence (PMS) low-mass stars embedded in the molecular cloud.

Orion is the nearest (d$\sim$414 pc) region of high-mass star formation, where three sites of  massive star formation have been identified: the Trapezium Cluster (TC), the Orion Hot Core (OHC), and OMC1-S. 
The TC contains the well-known four main-sequence (MS) massive Trapezium stars and is located in the central region of the optically visible Orion Nebula Cluster (ONC). The ONC has a size of $\sim$3 pc (Hillenbrand $\&$ Hartmann, \cite{hillenbrand-hartmann98}) and an age of about 10$^{6}$ yr (Hillenbrand, \cite{hillenbrand97}).
The OHC, located within the Orion KL nebula (Zuckerman et al., \cite{zuckerman81}; Shuping et al., \cite{shuping04}) is a very compact (size of $\sim$ 0.03 pc), hot (T$>$200K), and dense molecular condensation (mean density of n[H$_{2}$]=5$\times$10$^{7}$ cm$^{-3}$, Morris et al., \cite{morris80}), with a total luminosity of few 10$^{5}$ L$_{\odot}$ (de Vicente et al., \cite{devicente02}). The OHC is the prototype object for studying the formation of massive stars. Three massive objects with masses around 10 M$_{\odot}$ have been identified in this region: the Becklin-Neugebauer Object (BN), and sources {\em I} and {\em n}. 

OMC1-S ($\sim$10$^{4}$ L$_{\odot}$; Mezger et al., \cite{mezger90}; Rodr\'iguez-Franco et al., \cite{rodriguez-franco99a}), located $\sim$ 90\arcsec\ south of the OHC, is the other young massive star-forming region embedded in the Orion molecular cloud. OMC1-S  is a very active star forming region, harboring several outflows (Ziurys et al., \cite{ziurys90}; Rodr\'iguez-Franco et al., \cite{rodriguez-franco99a}, \cite{rodriguez-franco99b}; Zapata et al., \cite{zapata05}, \cite{zapata06}).

The Chandra Orion Ultradeep Project (COUP; Getman et al., \cite{getman05a}), based on a 10-day Chandra X-ray observation centered on the ONC, provided a unique data base of X-ray PMS stars. The X-ray emission from these low-mass stars can mainly be attributed to plasma heated to high temperatures by violent reconnection events in magnetic loops in the corona of PMS low-mass stars. Similar sources, variable in X-rays and in radio, have been detected in other low-mass star-forming regions with X-ray properties consistent with coronal magnetic activity (Feigelson \& Montmerle, \cite{feigelson99}; Forbrich et al., \cite{forbrich07}).  

In this paper, we extend previous works based on the COUP database to study the spatial distribution and estimate the densities of low-mass PMS stars. We find low-mass PMS stars clusters toward the three regions where massive star formation has recently occurred (TC), or it is currently taking place (OHC and OMC1-S). This fact and the high densities of PMS stars in these regions suggest that low-mass stars could play a role in the formation of massive stars.

\section{Data analysis}

 We analyzed the distribution and stellar densities of PMS low-mass stars with X-ray emission in Orion as a function of extinction. We used the Chandra COUP database provided by Getman et al. (\cite{getman05a}). We applied two different methods: i) spatial gridding, and ii) a close-neighbors method. This section is structured as follows: In 2.1. we present the two methods. Sections 2.2, 2.3, 2.4, and 2.5 are devoted to the effects introduced in our analysis by the cell size, the selected extinction ranges, the background/foreground contamination, and incompleteness.

\subsection{Methods}

\subsubsection{Spatial gridding}

We integrated the spatial distribution of the X-ray stars into a grid of square cells (see section 2.2. about the cell size), counting the number of stars contained in each cell. This method is very adequate to obtain maps of the stellar density, because it samples the entire region. However, spatial gridding can underestimate the real values of the stellar density peaks (see section 2.1.2.).

\subsubsection{Close-neighbors method}

The clustering of low-mass PMS stars may be affected by the spatial gridding used to compute the density of X-rays sources across the Orion region. For instance, depending on the grid origin, a stellar cluster could be within a single cell, or could be divided into two or more adjacent cells, leading to different values of the peak stellar densities. To estimate this effect we also analyzed of the distribution of close neighbors centered on every COUP source, which is independent of the grid origin. This method allows us to calculate the real stellar density peaks, but it is unable to generate density maps because it undersamples the field.

\subsection{Cell size}

From IR observations, Lada et al. (\cite{lada04}) found several dense stellar clusters in Orion, some of them with sizes of $<$ 0.05 pc. We are interested in detecting of these compact clusters, and in particular those connected with massive star formation. Massive stars are born in dense and hot condensations of gas, with typical sizes in the range 0.01 pc - 0.1 pc (Kurtz et al., \cite{kurtz00}), with a mean size of 0.03 pc (Saito et al., \cite{saito08}). This size also matches the size of the Orion Hot Core (Wright et al., \cite{wright96}). Therefore, we selected a cell size of 15\arcsec\ $\times$ 15\arcsec\ (0.03 pc$\times$0.03 pc) to match the mean size of the molecular cores where massive stars form as well as the sizes of dense stellar clusters. To evaluate the effect introduced by the cell size, we also repeated our analysis using a larger cell of 24\arcsec\ $\times$ 24\arcsec\ ($\sim$0.05 pc$\times$0.05 pc). We compare the results obtained with these two different cell sizes in section 3.2.1.

 \begin{figure*}
   \centering 
 \includegraphics[angle=0,width=15.5cm]{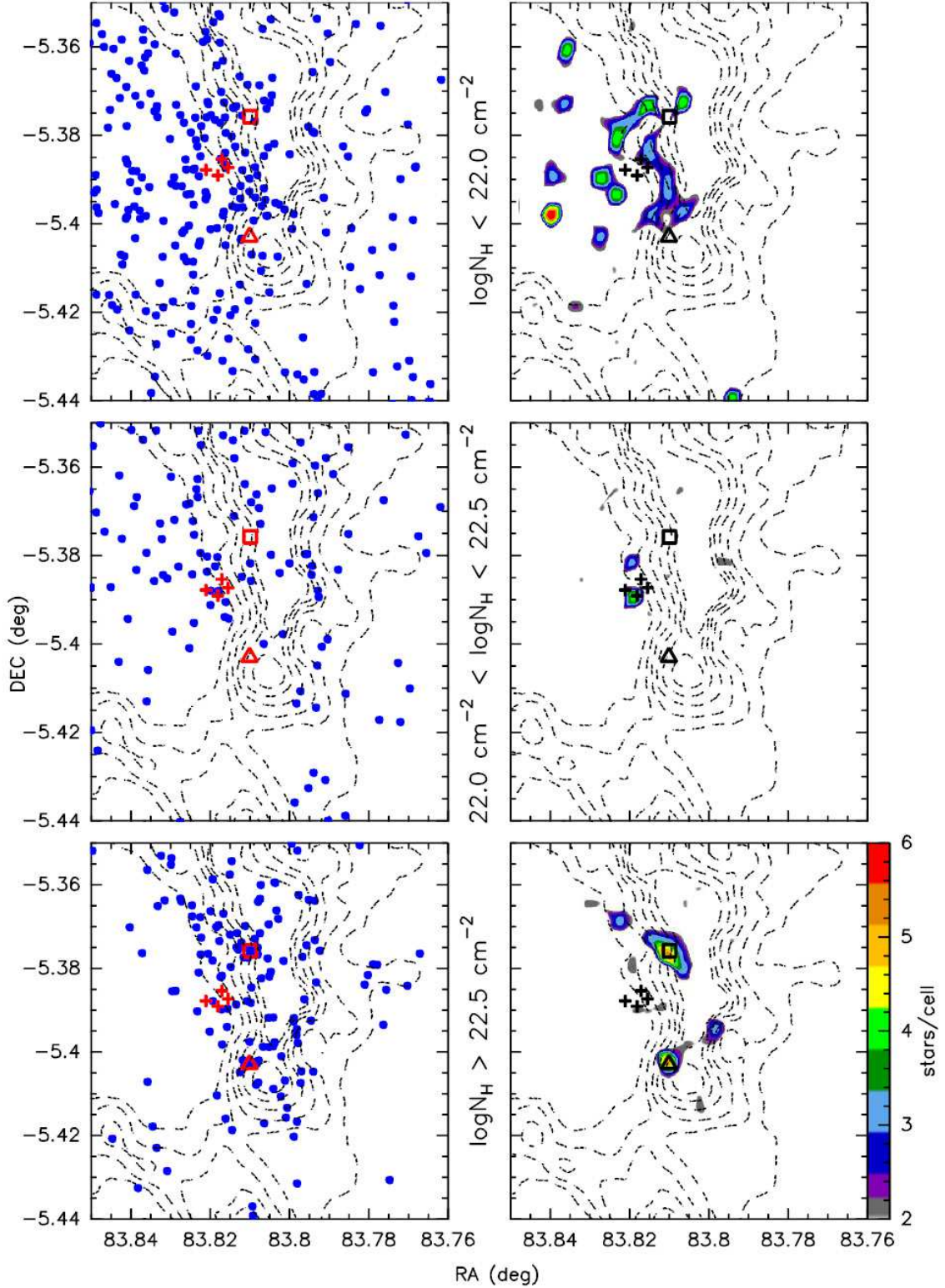}  
   \caption{Spatial distribution of low-mass stars for three different extinction ranges: logN$_{H}$$\textless$22.0 cm$^{-2}$ (upper panels), 22.0 cm$^{-2}$$\textless$logN$_{H}$$\textless$22.5 cm$^{-2}$ (middle panels), and logN$_{H}$$\textgreater$22.5 cm$^{-2}$ (lower panels). Dashed contours represent dense gas traced by the integrated intensity emission of 
CN (N=1-0) (Rodr\'iguez-Franco et al., \cite{rodriguez-franco}). The first contour level corresponds to 7 K km s$^{-1}$, the interval between contours is 4 K km s$^{-1}$. 
The left panels show the position of the X-ray stars as blue dots. Four crosses, the open square, and the open triangle show the location of the four main sequence massive Trapezium stars, the OHC, and the OMC1-S. The Right panels report the stellar surface density for the same extinction ranges, derived by counting the number of COUP sources using a cell size of 15\arcsec\ $\times$ 15\arcsec\ (0.03 pc $\times$ 0.03 pc) (right color scale), superimposed on the CN emission. The color contours have been smoothed for display purposes.}      
   \label{Fig1}
   \end{figure*}

 \begin{figure*}
   \centering
 \includegraphics[angle=0,width=18.5cm]{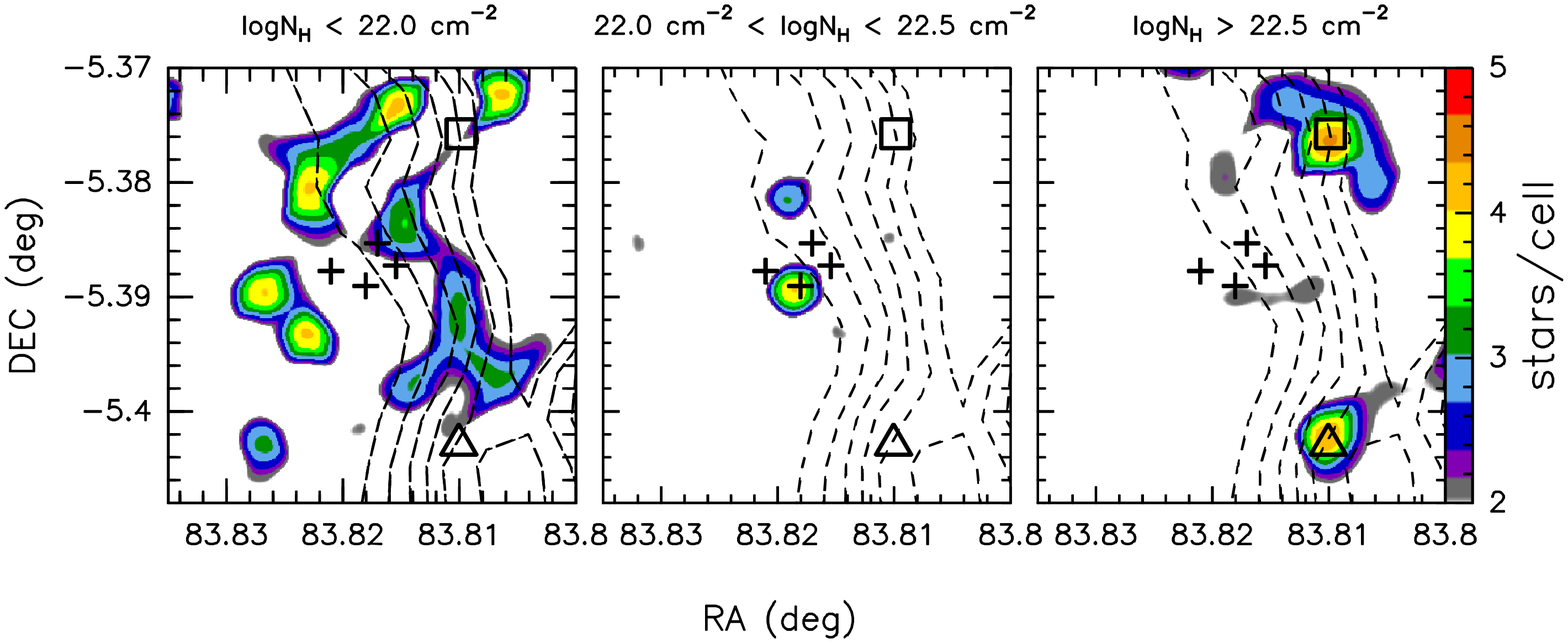}
 
   \caption{Zoom of the central region from Fig. 1. The color contours represent the stellar surface density (number of COUP sources using a cell box of 0.03 pc$\times$0.03 pc, right color scale) for the three extinction ranges (logN$_{H}$$\textless$22.0 cm$^{-2}$ in the left panel; 22.0 cm$^{-2}$$\textless$logN$_{H}$$\textless$22.5 cm$^{-2}$ in the middle panel; and logN$_{H}$$\textgreater$22.5 cm$^{-2}$ in the right panel). The color contours have been smoothed for display purposes. As in Fig. 1, the dashed contours represent the CN emission, and the four crosses, the open square, and the open triangle show the location of the four main-sequence massive Trapezium stars, the OHC, and the OMC1-S.}      
   \label{Fig2}
   \end{figure*}

\subsection{Selected extinction ranges}

Because the extinction across the Orion central region changes by several orders of magnitude (from 10$^{21}$ to 10$^{24}$ cm$^{-2}$), we analyzed the X-ray population as a function of three X-ray extinction ranges. The selection of the extinction bins are based on physical arguments. The first extinction range, logN$_{H}$$<$ 22.0 cm$^{-2}$ (A$_{V}$$<$5 mag), corresponds to the regions with low extinction that are mainly affected by the foreground gas in front of the optically visible ONC. 
 With the bin logN$_{H}$$>$22.5 cm$^{-2}$, corresponding to A$_{V}$$>$15 mag, we expect to find sources that are heavily embedded in dense regions of hot gas and dust (hot cores), where it is believed that massive protostars are found.
 The condition logN$_{H}$\textgreater 22.5 cm$^{-2}$ has also been used by Wang et al. (\cite{wang07}) to identify the heavily embedded population obscured by molecular material in the massive star-forming region NGC 6357. The population with 22.0 cm$^{-2}\textless$logN$_{H}$$\textless$22.5 cm$^{-2}$ (5$<$A$_{V}$$<$15 mag) corresponds to slightly extincted sources. In our analysis we only considered COUP sources with high enough X-ray counts to allow one to determine the value of logN$_{H}$ with an error of $<$0.5 dex.

\subsection{Background/foreground contamination}

One of the advantages of using X-rays to study massive star-forming regions is that they suffer less foreground- and background contamination than optical and infrared observations. Galactic contamination has very little impact on X-ray studies, but extragalactic sources can still be confused with embedded PMS stars. However, as Getman et al. (\cite{getman05b}) showed, the three regions studied in detail in this paper (i.e., the TC, OHC, and OMC1-S) are nearly free from contamination, so the detected sources reflect the actual cluster-member population. Despite this, we estimated a non-zero background/foreground level to provide a reasonable upper limit to the error of the stellar densities calculated in this work. A very conservative upper limit for the contamination would be to consider a uniform distribution of all COUP sources ($\sim$1600) in the field observed by Chandra (17\arcmin $\times$ 17\arcmin). This gives an error of $\pm$0.35 stars cell$^{-1}$, considering 0.03 pc$\times$0.03 pc cells. Getman et al. (\cite{getman05b}) claimed that only 217 COUP sources do not belong either to the ONC or the molecular cloud. Considering these sources, we obtain a lower limit for the background/foreground contamination of only 0.05 stars cell$^{-1}$. In our density analysis we consider a contamination level between the two values of 0.2 stars cell$^{-1}$, which is equivalent to 3 x 10$^{2}$ star pc$^{-2}$ in surface density and 2 x 10$^{4}$ stars pc$^{-3}$ in volume density.

\subsection{Incompleteness}

Although the COUP observations are quite deep and hence very sensitive, there is still a missing population of sources due to obscuration, especially in regions with high extinction, like the OHC and OMC1-S. By comparing the X-ray luminosity functions of the incomplete OHC and OMC1-S populations with the more complete ONC population, Grosso et al. (\cite{grosso05}) estimated that only about 48\% of the low-mass sources have been detected in the OHC, and 63$\%$ in the OMC1-S region. The TC is not as strongly obscured, and therefore its population is considered to be nearly complete.

\section{Results}

\subsection{Distribution of the low-mass stellar population}

 We used spatial gridding to obtain maps of the stellar density as a function of extinction. Fig. 1 shows the spatial distribution of X-ray low-mass stars for three different extinction ranges (see section 2.1.) superimposed on the total column density of dense molecular gas traced by the emission of CN (N=1-0) (dashed contours ; Rodr\'iguez-Franco et al., \cite{rodriguez-franco}). The left panels show the location of the X-ray stars as blue filled circles. Four crosses, an open square, and an open triangle indicate the location of the four TC stars, the OHC region, and the OMC1-S region. The right panels present the stellar surface density in terms of the number of COUP sources contained in cells of 15\arcsec\ $\times$ 15\arcsec\ (0.03 pc$\times$0.03 pc). A zoom of the central part of the field is presented in Fig. 2. Consistent with Grosso et al. (\cite{grosso05}), we found that PMS low-mass stars cluster toward the three regions related to massive stars: the TC, the OHC and the OMC1-S region.

The less extincted population, with logN$_{H}<$ 22.0 cm$^{-2}$ (upper panels in Fig. 1, left panel in Fig. 2), shows a ring-like structure around the TC. The slightly extincted stellar population, with 22.0 cm$^{-2}$$\textless$logN$_{H}$$\textless$22.5 cm$^{-2}$ (middle panels in Fig. 1 and 2), is clustered toward the position of the most massive star in the Trapezium, $\theta^{1}$ Ori C. 
Remarkably, this population partially fills the hole found in the non-embedded population. The most extincted X-ray sources (logN$_{H}$\textgreater 22.5 cm$^{-2}$; lower panels in Fig 1., right panel in Fig. 2), clearly associated with the molecular cloud traced by the CN emission, are clustered toward the OHC and the OMC1-S regions (open square and triangle, respectively, in Fig. 1 and 2), which host the ongoing massive star formation activity.

The selected extinction ranges have major implications on the outcome of this type of studies. Previous studies by Feigelson et al. (\cite{feigelson05}) and Grosso et al. (\cite{grosso05}) only considered two extinction ranges with logN$_{H}\textgreater$22.0 cm$^{-2}$ and logN$_{H}\textless$22.0 cm$^{-2}$.
Although these authors already reported the clustering toward the TC, OHC, and OMC1-S, their classification criteria did not allow them to clearly separate a non-extincted PMS population in the ring around the TC from the slightly extincted PMS stars toward the TC core, around the most MS massive star $\theta^{1}$ Ori C.

In addition, instead of the condition of logN$_{H}>$22.0 cm$^{-2}$ used by Feigelson et al. (\cite{feigelson05}) and Grosso et al. (\cite{grosso05}), we propose that the condition logN$_{H}$\textgreater 22.5 cm$^{-2}$ is more appropriate to isolate the most embedded population of low-mass PMS sources in the ongoing massive star-forming regions in Orion. Fig. 1 shows that the populations of sources with logN$_{H}<$22.5 cm$^{-2}$ (upper and middle panels) are distributed throughout the field, including the regions without molecular emission. However, the sources with logN$_{H}>$22.5 cm$^{-2}$ (lower panels) are predominantly concentrated in the region with CN molecular emission, suggesting their physical association with the molecular cloud.

\subsection{Stellar densities}

From spatial gridding  we can derive the stellar densities in terms of the number of stars per cell. However, as explained in section 2.1., this method may underestimate the values of the density peaks. For that reason, we used the close-neighbors method to calculate the densities in the low-mass PMS star clusters. In Fig. 3 we report the histograms of the number of 0.03 pc $\times$ 0.03 pc cells centered on COUP stars as a function of the number of stars found within each cell. The different colors and line types indicate the extinction ranges considered in our study. Consistently with our results from Fig. 1 and 2 (from spatial gridding), Fig. 3 (from the close-neighbors method) shows that the cells with the largest number of neighbors correspond to the X-ray sources detected toward the TC, the OHC, and the OMC1-S region independently of the spatial gridding. The densities obtained using cells centered on COUP stars are slightly higher than those calculated using the spatial gridding, and reflect the real value for the density peaks more reliably.

\begin{table*}
\tiny
\caption{Number of COUP stars in 15\arcsec $\times$ 15\arcsec (0.03 pc$\times$0.03 pc) cells centered on COUP sources (close-neighbors method), and derived stellar densities (surface and volume) for the density peaks seen at different extinctions.}             
\label{table:1}      
\centering          
\begin{tabular}{l l l l l}     
\hline\hline       
Region & Column density range & \# Sources &  Surface stellar density $\Sigma_{*}$  & Volume stellar density $\rho_{*}$ \\
 	     & (cm$^{-2}$) 		&  			&   (stars pc$^{-2}$) 					& (stars pc$^{-3}$) \\
\hline                    
   TC envelope  & $\textless$22.0 	& 5  & (5.5$\pm$ 0.3)$\cdot$10$^3$	& (3.5$\pm$ 0.2)$\cdot$10$^5$ \\ 
   TC core 		& 22.0-22.5 	    	& 4 & (4.4$\pm$ 0.3)$\cdot$10$^3$	& (2.8$\pm$ 0.2)$\cdot$10$^5$\\
   OMC1-S 	& $\textgreater$22.5 & 6 & (6.6$\pm$ 0.3)$\cdot$10$^3$	& (4.2$\pm$ 0.2)$\cdot$10$^5$\\
    OMC1-S (corrected)$^{a}$	& $\textgreater$22.5 & 10  & (1.10$\pm$ 0.03)$\cdot$10$^{4}$& (7.0$\pm$ 0.2)$\cdot$10$^{5}$\\
   OHC & $\textgreater$22.5 	& 9 & (1.10$\pm$ 0.03)$\cdot$10$^4$ & (7.0$\pm$ 0.2)$\cdot$10$^5$\\
   OHC (corrected)$^{a}$ 		& $\textgreater$22.5 &  19  & (2.10$\pm$ 0.03)$\cdot$10$^4$& (1.30$\pm$ 0.02)$\cdot$10$^6$\\ 
\hline                  
\end{tabular}
\begin{list}{}{}
\item[$^{\mathrm{a}}$]{Stellar densities corrected from incompleteness (see section 2.5).}
\end{list}
\end{table*}

The surface and volume stellar densities derived for the three extinction ranges toward the main density peaks (OHC, OMC1-S, the TC envelope, and the TC core) are summarized in Table 1. The surface densities were calculated by using 0.03 pc $\times$ 0.03 pc cells, and the volume densities using spheres with radius of 0.015$\,$pc, in both cases centered on COUP sources. The errors were estimated by considering $\pm$0.2 star cell$^{-1}$ as background/foreground contamination (section 2.4.).

The highest stellar density, with nine stars in the 0.03 pc cell in the range logN$_{H}$\textgreater 22.5 cm$^{-2}$, is found in two cells located within the OHC; one centered on COUP 621, likely a low-mass companion of the infrared source {\em n} (Lonsdale et al., \cite{lonsdale},  Grosso et al., \cite{grosso05}), and the other on COUP 622, located only 3.3\arcsec\ south of source {\em n}. 

These densities must be considered as lower limits since they do not consider the unseen population behind very high extinctions. This is particularly critical for the extinction range logN$_{H}$ \textgreater 22.5 cm$^{-2}$ (see section 2.5.). Considering the incompleteness due to obscuration, the surface and volume stellar densities in the OHC and OMC1-S could be as high as (7.0$\pm$ 0.2)$\times$10$^{5}$ stars pc$^{-3}$ and (1.30$\pm$ 0.02)$\times$10$^{6}$ for OMC1-S and the OHC, respectively. These densities may be even higher due to the presence of unresolved binaries. These values, along with the one found by Mart\'in-Pintado et al. (\cite{martin-pintado}) in Cep A HW2 (1.6$\times$10$^{8}$ stars pc$^{-3}$), are the highest found in all observed star clusters (Bally \& Zinnecker, \cite{bally05}).

In Fig. 4 we show the derived stellar surface density\footnote{We calculated the stellar densities counting the numbers of stars contained in rings with widths of $\sim$0.016 pc, centered on the cluster centers.} of PMS stars for the three extinction ranges versus the distance to the X-ray cluster centers. 
Left panel shows that the TC is dominated by slightly embedded sources concentrated in the central $\sim$ 0.03 pc, where the four MS massive stars of the Trapezium are located. The TC envelope formed by moderately extincted sources is located at a radius of $\sim$ 0.05 pc from the cluster center. The middle and right panels show that the embedded populations in the OHC and the OMC1-S PMS clusters are very compact with sizes $\sim$ 0.03 pc.

 \begin{figure}
   \centering 
   \includegraphics[angle=0,width=8.2cm]{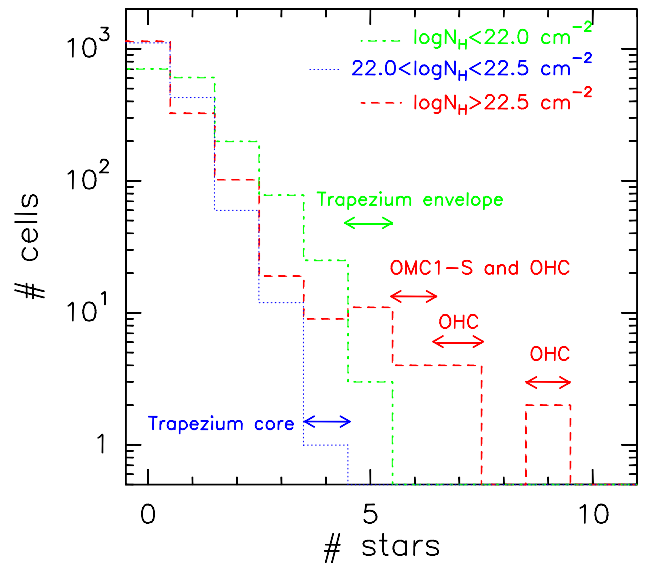}

      \caption{Number of cells versus the number of stars located inside the cell, considering 15\arcsec$\times$15\arcsec\ (0.03 pc$\times$0.03 pc) cells centered on each COUP X-ray star. Different colors and dashed lines show different extinction ranges. The densest cells (i.e., cells with the highest number of X-ray stars) are located in the OHC, OMC1-S, the TC core and the TC envelope (see text).}
         \label{Fig3}
 \end{figure}

\subsubsection{Effect of the cell size}

Table 2 compares the number of sources found in the main stellar density peaks (TC envelope, TC core, OMC1-S, and OHC) using the close neighbors method with different cells sizes (0.03 pc$\times$0.03 pc and 0.05 pc$\times$0.05 pc). We defined a {\it clustering parameter } ($\alpha$) as the ratio between the number of sources found in the 0.03 pc$\times$0.03 pc cell and the number found in the 0.05 pc $\times$ 0.05 pc cell. A value of $\alpha$ closer to 0 would mean that there is no central concentration of stars and hence that the distribution is not clustered. A value of $\alpha$=0.39 would correspond to a uniform stellar distribution. A value $\alpha$=1 would mean that all stars are concentrated in the smaller cell. Our results show that the clustering parameter $\alpha$ is between 0.7 and 0.8 in the clusters of PMS low-mass stars found in Orion (Table 2), which means that 70-80$\%$ of the stars in those clusters are concentrated in the inner cell of 0.03 pc. If we assume a Gaussian distribution for the stellar density $\rho_{*}$ $\propto$ e$^{-(2r/\gamma)^{2}}$ (where {\em r} is the distance to the center of the cluster and $\gamma$ is defined as the cluster size), $\alpha$=0.7-0.8 corresponds to $\gamma$ $\sim$ 0.024-0.028 pc. This size range is very similar to the size of our smaller cell. Consequently, our choice of the cell size (0.03 pc) is appropriate for sampling these clusters and for studying its stellar densities.

We note that the cell size is a key parameter for calculating stellar densities. The last column in Table 2 shows that the volume stellar densities calculated using 0.03 pc cells are $\sim$3 times higher than those calculated using larger cells of 0.05 pc. This is because the stellar density peaks are very concentrated ($\leq$0.03 pc). Therefore, the choice of an adequate cell size is the key in studying the stellar densities, in particular in the very young clusters of PMS low-mass stars, which are expected to be very compact.

\begin{table*}
\tiny
\caption{Comparison between the density analysis using the close-neighbors method, with cells centered on COUP sources with areas of 15\arcsec\ $\times$ 15\arcsec\ (0.03 pc$\times$0.03 pc) and 24\arcsec\ $\times$ 24\arcsec\ (0.05pc$\times$0.05pc).}             
\label{table:2}      
\centering          
\begin{tabular}{l l c c c c }   
\hline\hline       
Region & Column density range & \multicolumn{2}{c}{\# Sources} & Clustering parameter & Volume stellar density ratio  \\
 &(cm$^{-2}$) & 15\arcsec\ $\times$ 15\arcsec\ & 24\arcsec\ $\times$ 24\arcsec\ & $\alpha$ & $\rho_{*[0.03 pc]}$ / $\rho_{*[0.05 pc]}$ \\
 & & 0.03 pc $\times$0.03 pc  & 0.05 pc$\times$0.05 pc  & \\
\hline                    
   TC envelope  & $<$22.0 & 5  & 7& 0.7 & 2.9 \\ 
   TC core & 22.0-22.5 & 4 & 5 & 0.8 & 3.3 \\
   OMC1-S & $>$22.5 & 6 & 8 & 0.8 & 3.1\\
   OHC & $>$22.5 & 9 & 13 & 0.7 &2.8 \\ 
\hline                  
\end{tabular}
\end{table*}

 \begin{figure*}
   \centering 
   \includegraphics[angle=0,width=18.4cm]{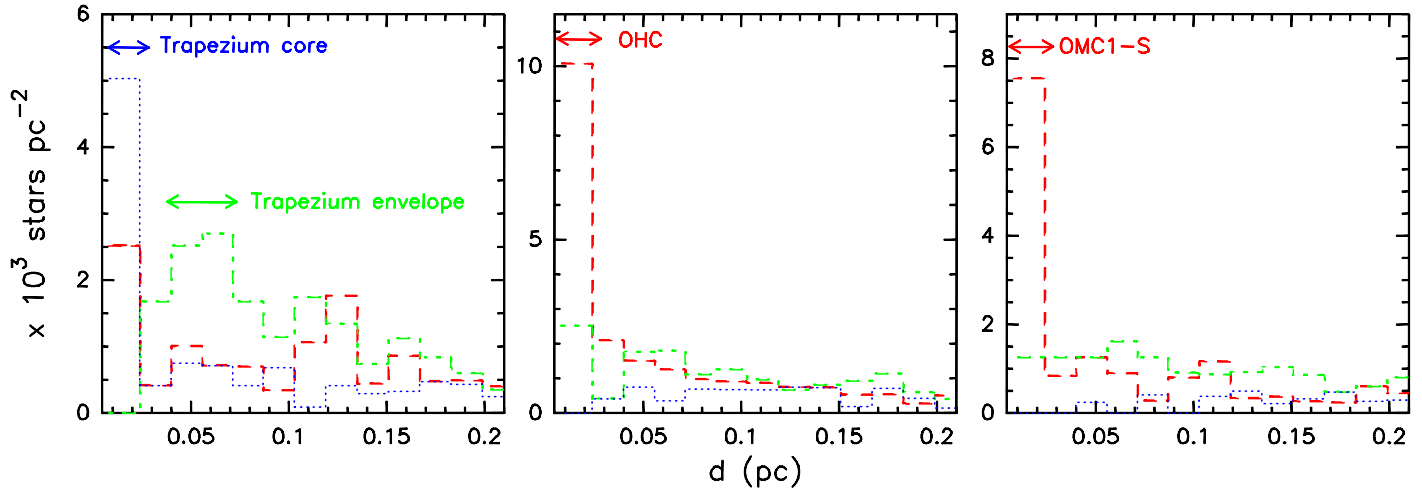}
      \caption{Surface stellar density versus the distance to the PMS cluster center for the TC (left), the OHC (middle), and OMC1-S (right). The different colors and dashed lines show different extinction ranges: green corresponds to logN$_{H}<$ 22.0 cm$^{-2}$, blue indicates an extinction range of 22.0 cm$^{-2}<$ logN$_{H}<$ 22.5 cm$^{-2}$, and red corresponds to logN$_{H}>$ 22.5 cm$^{-2}$.}
         \label{Fig4}
 \end{figure*}

\subsection{Comparison with other wavelengths}

We compared surveys of the stellar population in Orion carried out at different wavelengths with our analysis in X-rays. The different catalogs used for the comparison are presented in Table 3: one at optical (Hillenbrand, \cite{hillenbrand97}), and three at IR wavelengths (Hillenbrand \& Carpenter, \cite{hillenbrand-carpenter00};  Muench et al., \cite{muench02}; and Lada et al., \cite{lada04}). For the comparison we applied the same spatial gridding as for the X-rays, integrating the stellar population in cells of 0.03 pc $\times$ 0.03 pc. Fig. 5 shows the spatial distribution of the stars for the different surveys. Note that the color scale representing the number of sources per cell is different from that in Figs. 1 and 2.

 \begin{figure*}
   \centering 
   \includegraphics[angle=0,width=16.0cm]{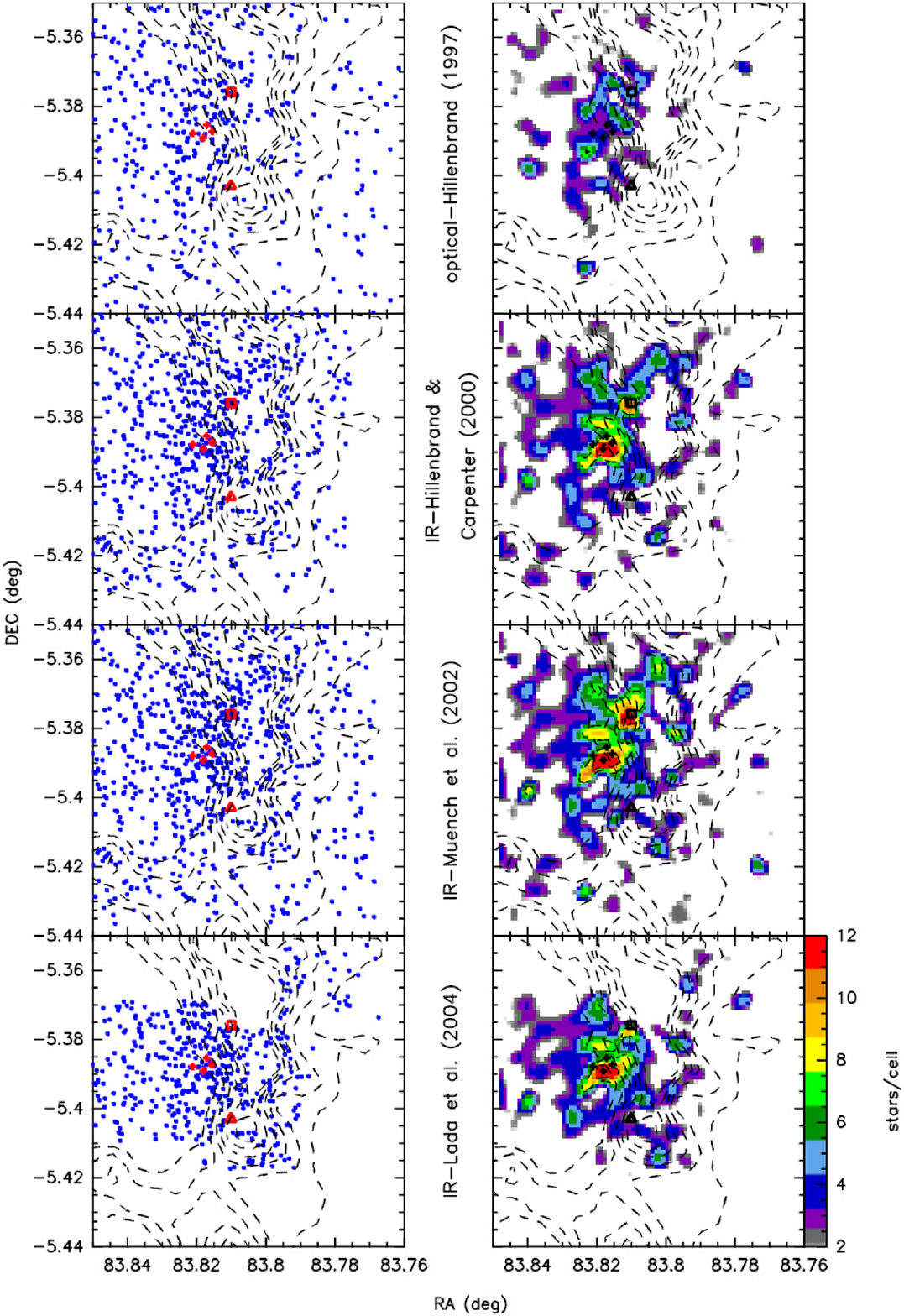}   
      \caption{Spatial distribution of stars from four different wavelengths. From upper to lower panels: optical catalog (V and I$_{C}$ bands) from Hillenbrand (\cite{hillenbrand97}), IR catalog (H and K bands) from Hillenbrand $\&$ Carpenter (\cite{hillenbrand-carpenter00}), IR catalog (J, H and K$_{S}$ bands) from Muench et al. (\cite{muench02}), and IR catalog (L' band) from Lada et al. (\cite{lada04}). As in Fig 1., dashed contours represent dense gas traced by the integrated intensity emission of CN (N=1-0) (Rodr\'iguez-Franco et al., \cite{rodriguez-franco}). The left panels show the position of the X-ray stars as blue dots. Four crosses, an open square and an open triangle show the location of the four main-sequence massive Trapezium stars, the OHC, and OMC1-S. The right panels report the stellar surface density for the same extinction ranges, derived by counting the number of COUP sources using a kernel box of 15\arcsec\ (0.03 pc)(right color scale), superimposed on the CN emission.}
         \label{Fig5}
 \end{figure*}

\subsubsection{Qualitative comparison with optical and IR observations}

We have shown in section 3.1. the importance of considering the population of stars in different extinction ranges, to distinguish members from the optically visible ONC and from the stellar population embedded in the molecular cloud located behind. In this subsection, however, we consider {\it all} stars observed at different wavelengths (Fig. 5), without applying any criteria to separate population at different extinctions. This prevents a reliable quantitative stellar density analysis study, because we are taking into account stars along the same line of sight that belong to different physical regions. However, this comparison between different wavelengths can give us interesting additional information.

The optical observation (upper panel in Fig. 5) reveals the ONC population, and presents a morphology consistent with the non- extincted X-ray sources (logN$_{H}<$22.0 cm$^{-2}$; upper panel in Fig. 1), with the main density peaks in the TC envelope (NE, NW and SE of the Trapezium stars), and without a significant population in the TC core. This gap in the radial distribution of low-mass stars surrounding the most massive star, $\theta^{1}$ Ori C, was already remarked by Hillenbrand (\cite{hillenbrand97}).

The three infrared observations show basically the same result. Unlike the optical population, the IR population is clustered in the TC core, around the most massive of the four Trapezium stars ($\theta^{1}$ Ori C) and in the OHC region. However, IR surveys do not detect the cluster found in X-rays toward OMC1-S. This is probably caused by a very high extinction toward this region.
 
The clustering in the TC core resembles the morphology presented by slightly extincted COUP sources (22.0 cm$^{-2}$$<$logN$_{H}$$<$22.5 cm$^{-2}$; middle panels in Fig 1). This, along with the fact that this clustering is observed at IR but not at optical wavelengths, suggests that the TC core contains more dust than its surroundings in the TC envelope, consistent with the X-ray results. This might yield some clues about the formation mechanism of the massive stars in the Trapezium (see the discussion in section 4.2.)

IR observations toward the OHC region detect both the ONC members located along the line of sight and the partially embedded sources  in the molecular cloud, which are still visible at these wavelengths. Rivilla et al. (\cite{rivilla13b}) found that stars in OHC with logN$_{H}$$<$22.9 cm$^{-2}$ (A$_{V}$$\sim$40 mag) are still visible in the IR, while those more embedded are, as expected, only visible in X-rays.

It would be expected that the L' band observations (Lada et al., \cite{lada04}; lower panel in Fig. 5), more sensitive to the presence of dust than IR observations at shorter wavelengths, show a clear clustering in the OHC. However, the peak in the OHC region is not as prominent as the peaks found in the other two surveys. This is because this region is located at the border of the Lada et al. (\cite{lada04}) observation (Fig. 5).

\subsubsection{Quantitative comparison with IR observations}

Similar to the procedure used to study the X-ray population in different extinction ranges, Lada et al. (\cite{lada04}) used the condition J-H$<$1.5 mag to select partially extincted and non-extincted sources (A$_{V}<$8 mag). These authors found that the main peak in the TC region (they called it Trapezium Core, but it is equivalent to the region containing our TC envelope and TC core) has a surface stellar density of 7.2$\times$10$^{3}$ stars pc$^{-2}$, a value of the same order (although slightly higher) as the average density of the TC envelope and TC core obtained from the X-ray data ($\sim$5$\times$10$^{3}$ stars pc$^{-2}$). 

Lada et al. (\cite{lada04}) also used the condition K-L$>$1.5 mag (roughly corresponding to A$_{V}>$15-26, and approximately equivalent to our condition logN$_{H}>$22.5 cm$^{-2}$) to select sources deeply embedded in molecular gas and dust. They found that the main peak in the embedded population is located in the OHC region, with a stellar density of 3.2$\times$10$^{3}$ stars pc$^{-2}$. This value is a factor 7 below the density we have calculated here (2.0$\times$10$^{4}$ stars pc$^{-2}$). This is because the most deeply embedded population remains unseen at IR wavelengths. Our work confirms the potential of X-ray observations to detect the population of heavily extincted low-mass stars in massive star cradles.

\subsection{The Trapezium Cluster: intermediate extincted core and non-extincted envelope}

 \begin{figure*}
   \centering 
   \includegraphics[angle=0,width=18cm]{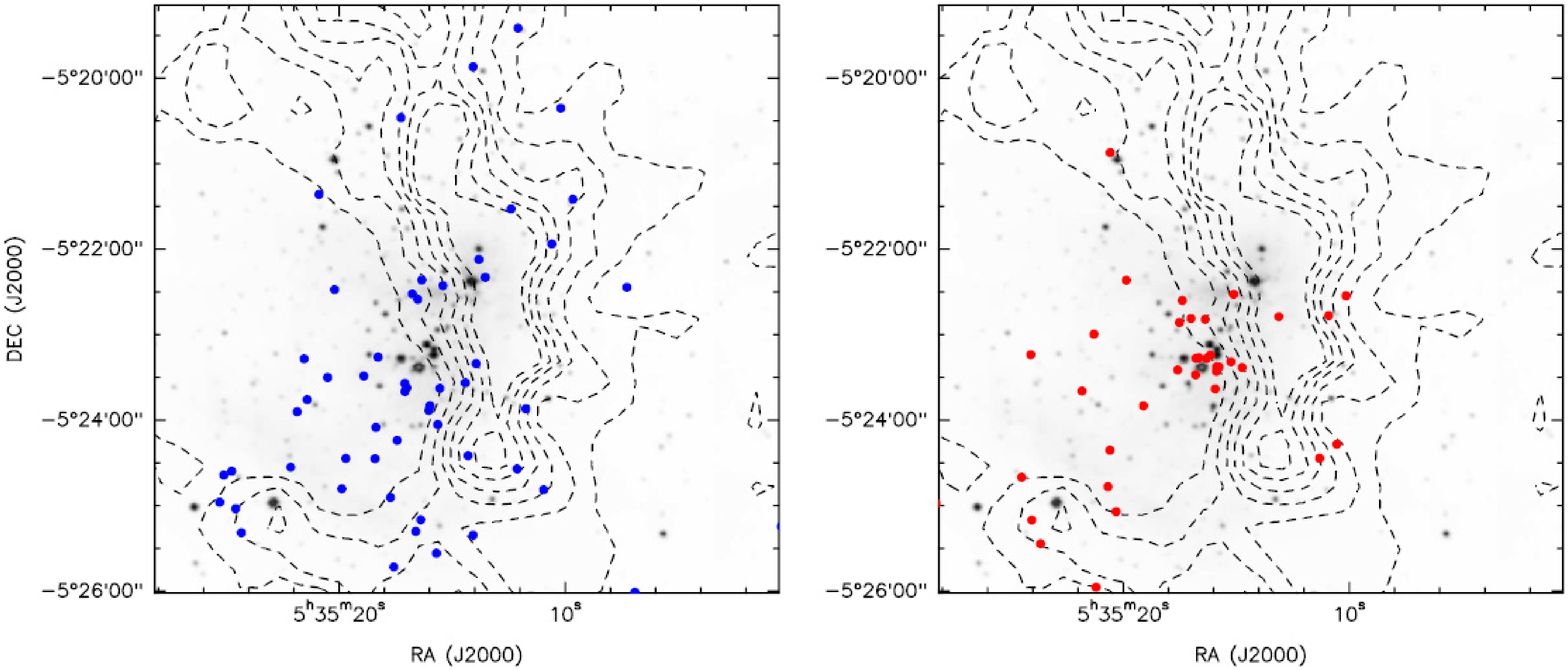}   
      \caption{Proplyds from the Kastner et al. (\cite{kastner05}) catalog with X-ray counterparts. The gray scale is the 2MASS image at K band, and the dashed contours denote the CN N=1-0 emission from the OMC (Rodr\'iguez-Franco et al., \cite{rodriguez-franco}) {\em Left}: stars with measured values of the hydrogen column density of logN$_{H}<$ 22.0 cm$^{-2}$. {\em Right:} stars with measured values of the hydrogen column density of logN$_{H}>$ 22.0 cm$^{-2}$.}
        \label{FigD1}
   \end{figure*}

Our analysis has revealed for the first time a dense cluster of X-ray low-mass stars with intermediate extinction (22.0 cm$^{-2}<$logN$_{H}<$22.5 cm$^{-2}$) around the most massive star in the Trapezium, $\theta^{1}$ Ori C, surrounded by a non-extincted envelope of stars. 
Most of these stars in the TC core were not detected in the optical survey from hillenbrand (\cite{hillenbrand97}), consistently with their intermediate extinction. We have shown that the gap found in the radial distribution in the optical survey disappears when considering intermediate extincted X-ray stars or IR stars.

To test whether the COUP stars clustered around $\theta^{1}$ Ori C are ONC members with extinctions higher than that usually found in the optical visible nebula, or whether they are simply slightly extincted members of the background OMC cloud, we crossed-checked the COUP catalog with optical HST observations of ONC proplyds from Kastner et al (\cite{kastner05}), which can detect stars more obscured than the optical Hillenbrand (\cite{hillenbrand97}) observations.
 
We found that in a close vicinity (20$\arcsec$ $\times$ 20$\arcsec$, $\sim$0.04 pc $\times$ 0.04 pc) of $\theta^{1}$ Ori C there is an absence of low extincted stars, and a cluster of seven X-ray stars with intermediate extinction, six of which are proplyds in Kastner et al. (\cite{kastner05}). This clearly confirms that these stars with intermediate extinction are ONC members.

In Fig. 6 we have plotted the spatial distribution of the proplyds with COUP counterparts, distinguishing those with logN$_{H}<$22.0 cm$^{-2}$ (blue dots in left panel) and logN$_{H}>$22.0 cm$^{-2}$ (red dots in right panel). The gray scale is the 2MASS image at K band, where the four massive Trapezium stars are clearly visible, and the dashed contours denote the CN N=1-0 emission from the OMC. Fig. 6 supports our finding of a core-envelope structure with intermediate extincted stars in the core and non-extincted stars in the envelope.

\begin{table*}
\caption{Surveys of Orion at different wavelengths.}             
\label{table:3}      
\centering          
\begin{tabular}{c c c }   
\hline\hline       
Wavelength & Telescope & Reference  \\
\hline     
\hline                 
   optical (bands V and I$_C$)  & KPNO 0.9 m + data from previous literature & Hillenbrand (\cite{hillenbrand97}) \\ 
 IR (H and K bands)& Keck I 10m & Hillenbrand $\&$ Carpenter (\cite{hillenbrand-carpenter00}) \\
IR (J, H and K$_{s}$ bands) & 1.2 m telescope at FLWO $ \& $ 3.5 m NTT in La Silla & Muench et al. (\cite{muench02}) \\
IR (L')& VLT UT1 & Lada et al. (\cite{lada04})\\
\hline                  
\end{tabular}
\end{table*}

\section{Discussion: Scenarios of massive star formation based on accretion theories and comparison with Orion observations}

Massive stars (M$>$8M$_{\odot}$) evolve very fast and start the nuclear burning when they are still accreting mass from the natal molecular core. The radiation pressure from the newly formed star has long been proposed to halt accretion, preventing the formation of the most massive stars (Wolfire $\&$ Cassinelli, \cite{wolfire87}). 

One of the first theories proposed to resolve this problem was the coalescence of low- and intermediate-mass stars in a dense stellar cluster (Bonnell et al., \cite{bonnell,bonnell02,bonnell05}), a violent process that does not require subsequent mass accretion. The observational evidence suggesting that a violent event occurred in the OHC region around 500 yr ago (Zapata et al., \cite{zapata09}, Rodr\'iguez et al,  G\'omez et al., \cite{gomez08}; Goddi et al., \cite{goddi11}), along with the high stellar density reported in our work ($\sim$ 10$^{6}$ stars pc$^{-3}$) would make coalescence possible in this region.
In Appendix A we discuss the possibility of a stellar collision in the high dense cluster revealed by Chandra in the OHC region, and conclude that a stellar interaction could have occurred in the cluster although it is uncertain whether this event has finally led to the formation of a massive star.
 
Stellar coalescence has always been seen as an exotic mechanism. This is because it requires higher stellar densities (10$^{6}$-10$^{9}$ stars pc$^{-3}$) than those usually found in stellar clusters to produce enough stellar collisions in 10$^{5}$-10$^{6}$ yr to form massive stars. Furthermore, this theory lost momentum when several works based on accretion showed that it was possible to overcome the radiation pressure problem without invoking coalescence (Yorke \& Sonhalter, \cite{yorke02}; Cesaroni et al.,  \cite{cesaroni06}). 

A general picture for massive star formation should likely be based on the two theories based on accretion: monolithic core accretion and competitive accretion. 
The main difference between these two theories is {\it how} the stars can accrete the mass needed to become massive.

According to the monolithic core accretion theory, initially proposed by McKee $\&$ Tan (\cite{mckee02,mckee03}), massive stars form from massive cores supported against self-gravity by its internal turbulence, with accretion rates high enough to overcome the radiation pressure. The massive core collapses monolithically and produces a single massive object or a few massive stars, rather than many low-mass stars (Krumholz et al., \cite{krumholz06,krumholz09}).  

The competitive accretion theory relies on the fact that nearly all massive stars are formed in stellar clusters with low-mass stars (Clarke, Bonnell $\&$ Hillenbrand, \cite{clarke00}; Lada \& Lada, \cite{lada03}). 
This theory assumes that the parental molecular cloud fragments into several condensations with masses around the Jeans mass\footnote{The Jeans mass, M$_{Jeans}$, is the mass necessary for an object to be gravitationally bound against its thermal support.}.
Numerical simulations have shown that gravitational fragmentation of a molecular cloud produces condensations with masses of about M$_{Jeans}$ $\leq$ 1 M$_{\odot}$ (Klessen et al., \cite{klessen98}; Bate et al., \cite{bate04}; Bonnell et al., \cite{bonnell04}) that collapse and form a low-mass star cluster. 
An additional mechanism is then needed to explain the formation of massive stars. It has been proposed that this mechanism is continuous gas accretion. Because fragmentation is inefficient in accumulating gas onto stars, there is a common gas reservoir in the core that can potentially be accreted (Bonnell et al., \cite{bonnell03}). Observations of pre-stellar structures and young clusters indeed show that most of the total mass (Krumholz $\&$ Bonnell, \cite{krumholz07}) is in a distributed gaseous form, so that it can potentially be accreted. Then, after low-mass stars are formed, they fall into local gravitational potential wells forming small-N clusters. This overall cluster potential well funnels gas down to the potential center, where it is captured via Bondi-Hoyle accretion by the "privileged" stars located at the potential center, progressively increasing their masses.

Clearly, core accretion and competitive accretion are rather different ways to explain massive star formation. However, it is not unreasonable to think of a scenario that involves a combination of the two theories, whenever the physical conditions allow them to work. Some authors have already suggested mixed scenarios for massive star formation (Peretto et al., \cite{peretto06}; Wang et al.,\cite{wang10}).

In this section we present three possible scenarios for massive star formation, where in principle both accretion theories could work. In practice, we show that the dominant mode of star formation is determined by the level of fragmentation in the core. While some works claim that radiation feedback from the stars can stop fragmentation (Krumholz et al., \cite{krumholz06}), and allows monolithic collapse to work, others claim that the core inevitably fragments into smaller condensations to form many stars (Clark $\&$ Bonnell, \cite{clark06}; Dobbs et al., \cite{dobbs05}; Bonnell \& Bate, \cite{bonnell06}; Peters et al., \cite{peters10}; Federrath et al., \cite{federrath10}). In this context, our scenarios consider different levels of fragmentation: i) low fragmentation and monolithic core accretion; ii) intermediate fragmentation and subcore accretion; iii) high fragmentation, subcore accretion and subsequent competitive accretion. 
We compare the predictions from each scenario with the results of our analysis presented in section 3.

 \begin{figure*}
   \centering  
   \includegraphics[angle=0,width=18cm]{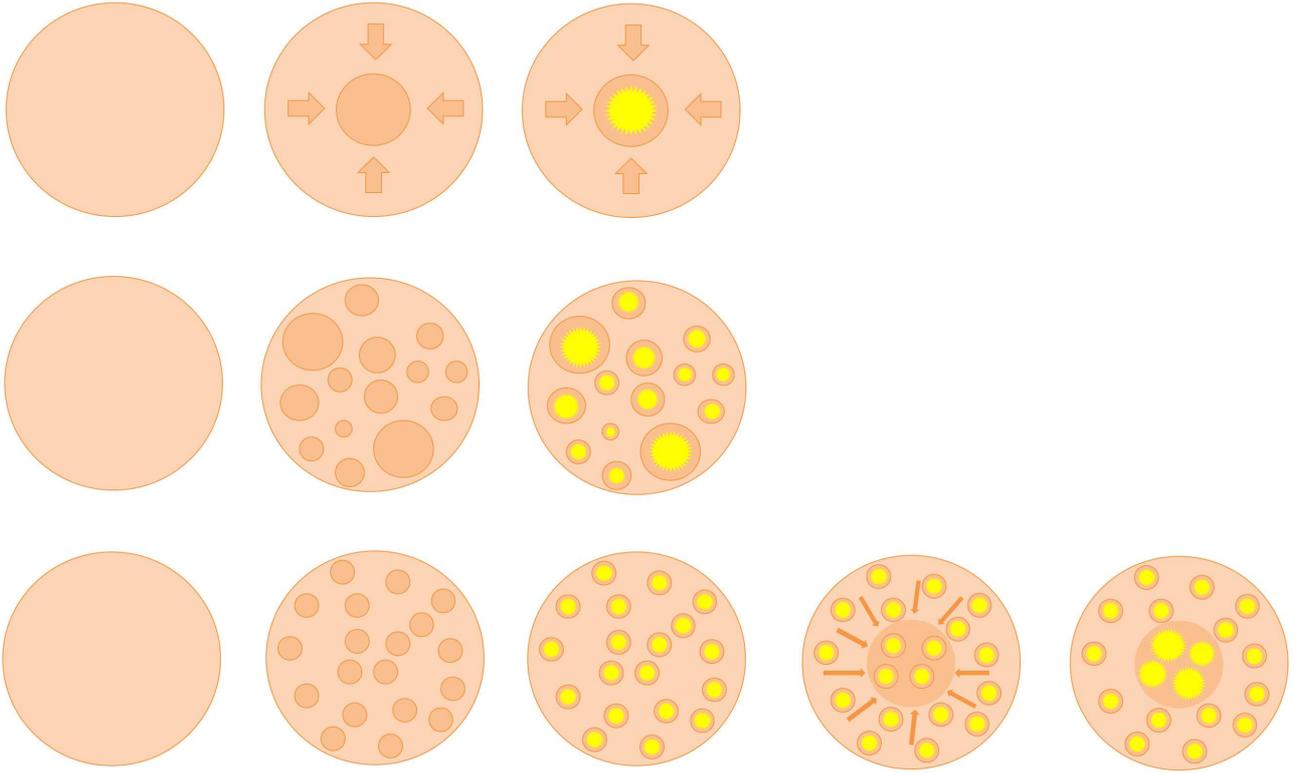}
      \caption{Scenarios for massive star formation based on the two accretion theories. The dominant role of each theory is mainly determined by the level of fragmentation in the core (see text). {\em Upper panels:} low fragmentation and monolithic core accretion. {\em Middle panel:} intermediate fragmentation and subsequent subcore accretion.  {\em Lower panel:} high fragmentation, subcore accretion forming stellar seeds, and competitive accretion.}
         \label{Fig6}
 \end{figure*}

\subsection{Low fragmentation: Monolithic core accretion}

In this scenario the radiative feedback halts fragmentation of the core, which collapses monolithically, allowing the formation of a single massive (or a few) objects. This is the core-to-star picture proposed by Krumholz (\cite{krumholz06,krumholz09}). This general collapse would rapidly consume the available gas, so that although subsequent Bondi-Hoyle accretion of the remanent gas could occur once the massive star is formed, the final mass is mainly determined by the core-accretion process. A simple schematic picture of this scenario is presented in the upper row in Fig. 7.
 
According to this scenario, massive stars should be born in isolation or be accompanied by other massive stars, rather than with a low-mass stellar cluster inside the core. Therefore, although the theory could explain the small fraction of massive stars found in isolation (de Wit et al., \cite{dewit04}), it would not explain the presence of low-mass stellar clusters with the massive star cradles in Orion (TC, the OHC and and OMC1-S; see section 3), and also in most stellar clusters (Clarke, Bonnell $\&$ Hillenbrand, \cite{clarke00}; Lada $\&$ Lada, \cite{lada03}).

In addition, it is known that massive cores in infrared dark clouds (IRDCs), believed to represent the initial conditions for massive stars and star cluster formation, are usually found to fragment into smaller condensations of low- to intermediate masses (see e.g. Wang et al., \cite{wang11}), which is in contrast with the idea of a massive star being formed from a single massive core. Other works (Bonnell $\&$ Bate, \cite{bonnell05}; Dobbs et al., \cite{dobbs05}) have suggested that such a massive core, even if supported by turbulence, very likely fragments to form many stars.

\subsection{Intermediate fragmentation: subcore accretion}

One could think of an scenario based on monolithic collapse, but acting at subcore scales. In this scenario, the core would suffer fragmentation, generating several subcores with different masses and sizes. The subcores with lower masses would produce low-mass stars, while those with higher masses could potentially form massive stars, in both cases via subcore accretion. A schematic picture explaining this scenario is presented in the middle row in Fig. 7. In principle, massive stars could be formed in isolation or be accompanied by low-mass stars, depending on how the core initially fragments into subcores.

Once the stellar cluster is born, subsequent Bondi-Hoyle accretion from the gas reservoir still available in the core would be possible. However, it is difficult for competitive accretion to work in this context because there is no overall well-defined central cluster potential well, but a collection of local potential wells around each formed star. Therefore, there is no efficient funneling of gas to the center of the stellar cluster.

This scenario is very similar to the results presented by the simulations by Krumholz et al. (\cite{krumholz11,krumholz12}) and it is able to produce stars with different masses. These works implement the theory of core accretion in the general collapse of a cloud instead of the single massive core considered in previous simulations (Krumholz et al., \cite{krumholz06,krumholz09}). The outcome of their simulations includes some massive stars formed nearly in isolation (consistently with their previous works) and also others located in low-mass stellar clusters, which would be more compatible with our results in Orion.

However, this scenario presents several problems. Krumholz et al. (\cite{krumholz11,krumholz12}) argued that radiative feedback from the stars stops further fragmentation, explaining why not all subcores reach low-masses $\sim$ 1 M$_{\odot}$ during the collapse phase. However, it is not clear that radiative feedback from the newly stars could be so efficient to stop fragmentation. Several other works point in the opposite direction, i.e., massive cores should fragment into many stars (Dobbs et al., \cite{dobbs05};  Clark \& Bonnell, \cite{clark06}; Bonnell \& Bate, \cite{bonnell06}; Federrath et al., \cite{federrath10}; Peters et al., \cite{peters10}), especially with the presence of outflows that are not included in the simulation by Krumholz et al. (\cite{krumholz11,krumholz12}) and could favor the fragmentation (Cunningham et al., \cite{cunningham11}).

In addition, there is other evidence indicating that this scenario does not fully explain, at least, the massive star formation that occurred in Orion. In this scenario fragmentation does not always produce more massive subcores at the center of the initial core (see Krumholz et al. \cite{krumholz11,krumholz12}). Therefore, one would not expect a trend for massive stars to form at the center of the core, as seen in the TC (see discussion in section 4.3).

\subsection{High fragmentation: subcore accretion forming stellar seeds and subsequent competitive accretion}

We assume now that the core fragments into subcores, which collapse producing a cluster of low-mass stars via subcore accretion. Only a fraction of the gas is directly incorporated into the low-mass stars, while the rest is distributed throughout the core. In this context, the stars compete to accrete this material. The stars located near the center of the potential well created by the whole cluster benefit from the higher gravitational attraction and gather matter via Bondi-Hoyle accretion at much higher rates, becoming higher mass stars. The stars that are not at the center of the potential well do not accrete significant amounts of gas, and remain  low- and intermediate-mass stars. This scenario with high fragmentation (lower panel in Fig. 7) is essentially coincident with competitive accretion, and naturally produces a $whole$ cluster of stars with different masses.
 
Bonnell et al. (\cite{bonnell04}) simulated the evolution of a turbulent molecular cloud with properties similar to those in Orion, where fragmentation and competitive accretion play the main role in the formation of massive stars. The result of their simulation (see their Figure 5 and also Fig. 14 in Zinnecker $\&$ Yorke, \cite{zinnecker07}) shows a $\sim$ 0.6 pc $\times$ 0.6 pc region where the massive stars are formed in the center of low-mass star clusters. The similarity between the morphology of the whole cluster with the Orion observations presented in this paper is remarkable (Fig. 1). Moreover, there is a trend in the TC for higher mass stars to be found at closer distances to the most massive star $\theta^{1}$ Ori C. Fig. 8 shows the mass of the stars in the TC versus the distance to  $\theta^{1}$ Ori C (data from the optical observations by Hillenbrand, \cite{hillenbrand97}). This mass segregation is primordial, i.e., the massive stars were initially born in the cluster center without migration (Bonnell, Larson $\&$ Zinnecker, \cite{bonnell07}; Reggiani et al., \cite{reggiani11}). Therefore, this suggests a preference for the most massive stars to be born in the densest and central parts of the low-mass star cluster, as expected in the competitive accretion theory.

This scenario can also naturally explains the structure of extincted and non-extincted low-mass stars found in the TC (Fig. 1, 2 and 6).
The non-extincted stars in the envelope of the TC could have lost the battle of competitive accretion - keeping their low-mass - against the four Trapezium stars in the core, which would have grown until they reached their current high masses. 
The non-extincted PMS stars in the envelope of the TC are therefore devoid of gas while those in the core are still partially embedded in gas and dust.

Given that this scenario provides the best explanation for our results, we evaluated whether Bondi-Hoyle accretion can indeed gather enough mass to produce a massive star in the Orion environment. The accretion rate $\dot{m}$ is (Bonnell \& Bate, \cite{bonnell06})

\begin{equation}
\centering
\dot{m} = 4\pi\rho  \frac{(GM)^2}{v^3} , 
\end{equation}

where $\rho$ is the volume gas density, G the gravitational constant, and $v$ the relative velocity between the gas and the accreting star. Considering that the relative gas velocity follows the turbulent velocity size-scale relation  $v\sim R^{1/2}$ (Larson et al., \cite{larson81}; Heyer $\&$ Brunt, \cite{heyer04}), and assuming that the gas density $\rho$ is centrally condensed with $\rho\sim R^{-1}$ (as found in other massive star-forming regions like Monoceros R2, Choi et al., \cite{choi00}), the relation between the velocity and the mass of the core is $v\sim M^{1/4}$. Combining this with eq. 1 and integrating this expression, one can obtain the time needed to build up a massive star

 \begin{equation}
\centering
t=t_{seed}+\frac{v^{3}}{\pi\rho G^{2} M_{i}^{3/4}}\left(\frac{1}{M_{i}^{1/4}}-\frac{1}{M^{1/4}}\right) ,
   \end{equation}

where $M$ is the mass of the star and $M_{i}$ is the initial mass of the "stellar seed", and t$_{seed}$ is the time needed to form the stellar seed with initial mass M$_{i}$.

The relative velocity of the gas $v$ is a key parameter to determine whether competitive accretion is efficient or not (Krumholz et al., \cite{krumholz05}; Bonnell $\&$ Bate, \cite{bonnell06}).
It has been suggested that turbulence injected by winds, protostellar outflows, or ionized regions could increase the relative velocity of the gas, preventing competitive accretion from being the dominant process in the massive star formation (Krumholz et al., \cite{krumholz05}). However, the results presented in this work have shown that massive stars in Orion have formed in small-N clusters (tens of stars, see Table 1), where the gas velocity dispersion is expected to be lower (Bonnell et al, \cite{bonnell03}, \cite{bonnell06}).

 \begin{figure}
   \centering  
  \includegraphics[angle=0,width=8cm]{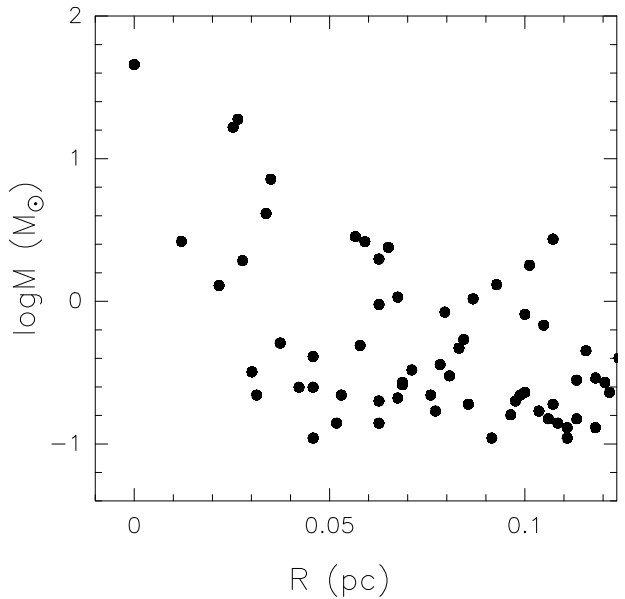}
      \caption{Mass of the stars in the TC versus the distance to the most massive star $\theta^{1}$ Ori C (data from the optical observation presented by Hillenbrand, \cite{hillenbrand97}).}
         \label{Fig8}
 \end{figure}
 
Assuming a Maxwellian distribution of velocities, we used the root mean square of the gas velocity, $v_{RMS}$, as a good estimate of the relative velocity $v$. The value of $v_{RMS}$ is related to the linewidth of the molecular lines, $\Delta v$, with the expression

\begin{equation}
\centering
v_{RMS}=\frac{1}{2}\left(\sqrt{\frac{3}{2\,Ln2}}\right)\Delta v ,
\end{equation}

We use here the linewidth of vibrationally excited HC$_3$N (HC$_3$N*, hereafter) because this molecule is particularly well suited to study the kinematics of the dense gas close to very young massive stars (Wyrowski et al., \cite{wyrowski99}; de Vicente et al., \cite{devicente00,devicente02}). 
The OHC observations by de Vicente et al. (\cite{devicente02}) and Rivilla et al. (\cite{rivilla13c}, in preparation) show that $\Delta v\sim $ 7 km s$^{-1}$. This value is consistent with those observed with other dense molecular tracers as NH$_{3}$ ($\sim$5 km $^{-1}$, Goddi et al., \cite{goddi11b}) or CH$_{3}$OH (4-6 km s$^{-1}$, Peng et al. \cite{peng12}). Using eq. 3, the HC$_3$N* linewidth is equivalent to v$_{RMS}$=5.2 km s$^{-1}$.

Fig. 9 shows the time needed to form a massive star from stellar seeds with M$_{i}$=1 M$_{\odot}$ by subcore accretion and subsequent competitive accretion. The blue line indicates the accretion time from the expressions from Krumholz et al., (\cite{krumholz12}) and McKee $\&$ Tan (\cite{mckee03}) to form the stellar seed. The red line corresponds to the Bondy-Hoyle accretion time using eq. 2, with $v=v_{RMS}$=5.2 km s$^{-1}$. The volume- and surface densities used are $\rho$=1.95$\times$10$^{-16}$ g cm$^{-3}$ and $\Sigma$=12 g cm$^{-2}$, calculated from the average gas density in the OHC of N[H$_{2}$]=5$\times$10$^{7}$ cm$^{-3}$ (Morris et al., \cite{morris80}). According to Fig. 9, a 10 M$_{\odot}$ star (approximately the mass estimated for the massive objects found in this region) can be formed in the OHC by accretion of gas in $<$ 4$\times$10$^{5}$ years. 
We note that this accretion time estimate does not demonstrate by itself that massive stars are formed by competitive accretion in the OHC, but it shows that
it could be possible.

In the OMC1-S region, where the gas density is lower but on the order of the OHC region (see CN contours in Fig. 1 and 2), and where a similar (or lower) turbulence may be expected, massive stars could form in the same way.

In the case of the TC, the current gas density is $\sim$10$^{-17}$ g cm$^{-3}$ (Hillenbrand $\&$ Hartmann, \cite{hillenbrand-hartmann98} ; Bonnell $\&$ Bate, \cite{bonnell06}), one order of magnitude lower than that in the OHC ($\sim$1.95$\times$10$^{-16}$ g cm$^{-3}$). This implies that competitive accretion could be less important according to eq. 1. However, as indicated by Krumholz et al. (\cite{krumholz11}), this density must have been higher by an order of magnitude at the moment where the stars were formed, because the cluster has likely expanded due to gas ejection (Kroupa et al., \cite{kroupa01}; Tan et al., \cite{tan06}). Then, competitive accretion could still have been responsible for the formation of the massive stars in the Trapezium.

  \begin{figure}
   \centering  
   \includegraphics[angle=0,width=8cm]{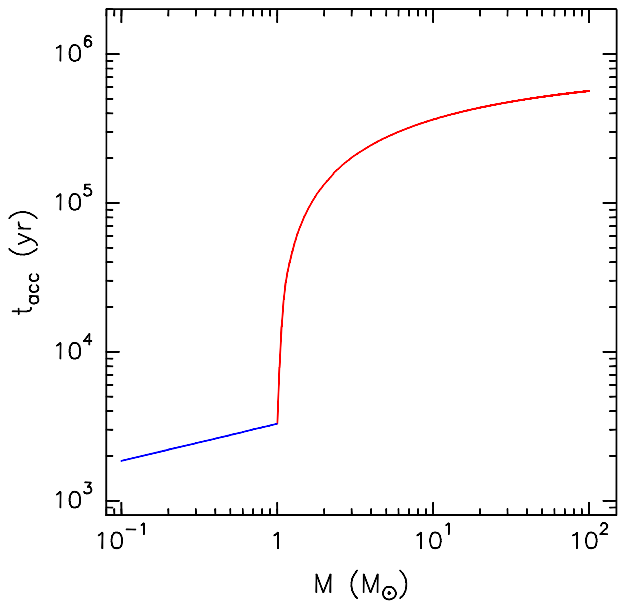}
      \caption{Accretion time necessary to form a massive star in the OHC in the high fragmentation scenario. The blue line corresponds to the time needed to form the stellar seeds of 1 M$_{\odot}$ via subcore accretion. We have used the expressions from Krumholz et al. (\cite{krumholz12}) for the accretion rate. The red line corresponds to the accretion time needed to reach the final mass M, using eq. (2). The volume- and surface densities used are $\rho$=1.95$\times$10$^{-16}$ g cm$^{-3}$ and $\Sigma$=12 g cm$^{-2}$.}
      \label{Fig8}
 \end{figure}

In summary, from our combined data (CN molecular emission and X-ray PMS stars), it seems that at least two basic ingredients are needed to form massive stars: i) dense gas, and ii) a cluster of low-mass stars. It is remarkable that the low-mass star clusters appear only at certain positions of the north-south elongated ridge of dense molecular material (Fig. 1), suggesting that dense molecular material only does not necessarily form massive stars.  This idea has also been suggested by Krumholz et al. (\cite{krumholz12}). Along with dense gas, our results suggest that the presence of low-mass star clusters is also a necessary but not sufficient condition for massive star formation. This is supported by the existence of dense low-mass star clusters, like the one at $\sim$[83.84, -5.395] (see Fig. 1), which does not present any massive star. This low-mass star cluster is seen in the non-extincted column density range (logN$_{H}<$22.0 cm$^{-2}$) in the X-ray observations (upper panel in Fig. 1), and also in the IR surveys and in the optical observation by Hillenbrand (\cite{hillenbrand97}; see upper panel in Fig. 5), confirming that it belongs to the ONC. In this region, the lack of dense molecular gas suggests that this low-mass star cluster was unable to accrete additional material to increase its mass to form more massive objects.

\section{Conclusions}

We have analyzed the distribution of PMS low-mass stars with X-ray emission in Orion observed by Chandra as a function of extinction, and calculated their stellar densities. Our results show that PMS low-mass stars cluster toward the three regions connected to massive stars: the Trapezium Cluster (TC), the Orion Hot Core (OHC), and the OMC1-S region. We derived PMS low-mass stellar densities of 10$^{5}$ stars pc$^{-3}$ in the TC and OMC1-S, and 10$^{6}$ stars pc$^{-3}$ in the OHC. These very high stellar densities and the close association between the low-mass star clusters with the regions where massive star formation has recently occurred (TC) or it is currently taking place (OHC and OMC1-S) suggest that not only dense gas is needed to form massive objects, but also a cluster of low-mass stars. 

The X-ray observations show for the first time in the TC that low-mass stars with intermediate extinction are clustered toward the position of the most massive star $\theta^{1}$ Ori C, which is surrounded by a ring of non-extincted PMS low-mass stars. This 'envelope-core' structure is also supported from infrared and optical observations.

We proposed several scenarios for massive star formation in which the theories based on accretion (core accretion and competitive accretion) play different roles and the parental core presents different levels of fragmentation: i) low fragmentation and monolithic core accretion; ii) intermediate fragmentation and subcore accretion; iii) high fragmentation, subcore accretion and subsequent competitive accretion. We compared the predictions of each scenario with the findings of our analysis of the X-ray observations by COUP.

The first scenario, based on the classic core accretion theory (core-to-star), does not explain the formation of a significant population of low-mass stars, contradicting the results presented in this paper. 

We showed that the second scenario (intermediate fragmentation) is not fully appropriate to explain our results. In this scenario fragmentation does not always produce more massive subcores at the center of the parental core. Therefore, we would not expect the trend for massive stars to form at the center of the core that we observe in the TC. Moreover, the presence of a significant low-mass stellar population revealed by X-rays in the massive star cradles suggests that the level of fragmentation of the initial core is high. 

This led us to a third scenario, with higher fragmentation forming a low-mass stellar cluster via subcore accretion. These $^{\prime\prime}$stellar seeds$^{\prime\prime}$ can gather enough mass by competitive accretion, forming more massive stars at the cluster center. Therefore, the outcome of this scenario predicts our findings. Furthermore, we showed that the formation of massive stars in a core with the physical conditions measured in the OHC can be explained in terms of competitive accretion. Competitive accretion could also work in OMC1-S, with similar gas properties, and in the TC, which was denser at the moment when the stars were formed. 

Finally, given that the stellar density in the OHC is one of the highest reported in the galaxy (1.30$\times$10$^{6}$ stars pc$^{-3}$) and because there is much observational evidence that a violent process occurred in the core, we discuss the possibility of a coalescence event in the OHC stellar cluster in Appendix A. We conclude that although coalescence does not seem a common mechanism for building up massive stars, a single stellar merger may occur in the evolution of the cluster of the OHC, favored for the presence of disks, binaries, and accretion.

\begin{acknowledgements}
We thank the Spanish MICINN for the support provided through grants ESP2007-65812-C02-C01 and AYA2010-21697-C05-01, and AstroMadrid (CAM S2009/ESP-1496), and the Spanish CSIC for the grant Jae-Predoc2008. V.M.R. acknowledges Jorge Sanz-Forcada for the very useful discussions about the X-ray science contained in this work.
\end{acknowledgements}

\appendix

\section{Stellar collisions in dense clusters}

The coalescence of low- and intermediate-mass stars has been proposed as one of the mechanisms that may form massive stars (Bonnell et al., \cite{bonnell}; Stahler et al., \cite{stahler00}; Bonnell \& Bate, \cite{bonnell02}; Bally \& Zinnecker, \cite{bally05}). Observations have shown that stars are almost never born in isolation but favor formation in clusters (Lada \& Lada 2003). Bonnell et al. (\cite{bonnell}) and Stahler et al. (\cite{stahler00}) argued that high stellar densities may exist for a brief period of time during the very embedded early evolution of stellar clusters in molecular cores, making stellar collisions more likely. Therefore, interactions could occurr among members of a protostellar cluster, especially at the early stages of cluster evolution. These interactions could play a role in the formation of high-mass stars.

Coalescence requires stellar densities in the order of 10$^{6}$-10$^{8}$ stars pc$^{-3}$ (Bonnell \& Bate, \cite{bonnell05}; Bally \& Zinnecker, \cite{bally05}) that are higher than the highest densities (of $\sim$ 10$^{5}$ stars pc$^{-3}$) found observationally (e.g. Claussen et al., \cite{claussen94} or Figer et al., \cite{figer02}). 
However, we have detected in this work a very high stellar density in the OHC region of 10$^{6}$ stars pc$^{-3}$, a range where stellar collisions may occur. 
This indicates that a collision could be possible in the OHC. This idea is also supported by some observational evidence that suggests that a violent event occurred in the OHC in the past: a large-scale CO outflow (Kwan \& Scoville, \cite{kwan76}), the H$_2$ fingers (Taylor et al., \cite{taylor84}; Allen \& Burton, \cite{allen93}),  the CO filaments (Zapata et al., \cite{zapata09}), and the proper motions measured for the massive objects BN, I and n (Rodr\'iguez et al, \cite{rodriguez05}; G\'omez et al., \cite{gomez08}; Goddi et al., \cite{goddi11}), all of them pointing toward approximately the same central position within the OHC region (see Fig. 1 of Zapata et al., \cite{zapata11}).
 Rodr\'iguez et al. (\cite{rodriguez05}) and Zapata et al. (\cite{zapata09}) suggested that this violent event could have been triggered by an interaction among several stars in the protostellar cluster. This interaction is consistent with a close encounter that would induce stars to coalescence in this particular high-density cluster revealed by X-ray observations.

In this appendix, we derive some expressions to describe stellar collisions in a cluster and apply them to the properties of the OHC with the aim to determine if a stellar collision was possible in the region. We also compare our results with other works that include dynamical simulations of stellar encounters.

\subsection{Physics of stellar collisions}

We follow Binney \& Tremaine \cite{binney87} to study the case where two naked stars encounter (A.1.1) and extend this case to a encounter between a star and a star with a circumstellar disk (A.1.2). In section A.1.3 we present the possibility of disk-induced stellar capture in an encounter forming a binary, and finally in A.1.4, we give a simple analysis of the orbital decay of a binary trigger by accretion onto the system, which may produce a collision between the two stars.

\subsubsection{Direct star-star encounter}

We consider an encounter between two stars with mass M. Applying conservation of momentum and energy in the system, it is possible to obtain the collision rate (essentially the inverse of the time needed to produce a collision, $\tau_{coll}$)

 \begin{equation}
\centering
      \frac{1}{\tau_{coll}}$ \, $[yr^{-1}]=4\sqrt{\pi}\rho_{*}\sigma_{*}r_{coll}^{2}\left(1+\frac{GM}{\sigma_{*}^{2}r_{coll}}\right) ,
   \end{equation}

where $\rho_*$ is the volume stellar density and $\sigma_{*}$ is the velocity dispersion of the cluster. The first term corresponds to the geometrical cross section of the stars, and the second is due to gravitational focusing. The two stars collide when the collision distance r$_{coll}$ is equal to the distance between the centers of mass of the stars, i.e., the sum of their radii: r$_{coll}$ = R$_{1}$ + R$_{2}$.

\subsubsection{Star-(star+disk) encounter}

In this section, we extend the previous case by adding the presence of a circumstellar disk. Calculating the gravitational potential generated by a disk is difficult in general (Binney \& Tremaine, \cite{binney87}). For this reason, we made a reasonable approximation to characterize the disk, so that the calculation of its potential is rather simple. We consider the disk to be an infinitesimal thin surface, with surface density

\begin{equation}
\centering
\Sigma(r)=\beta\frac{r_{d}M_{d}}{2\pi\left(r^{2}+r_{d}^{2}\right)^{3/2}} ,   
\end{equation}

where $\beta=(1-1/\sqrt{2})^{-1}$, r$_{d}$ and M$_{d}$ are the radius and mass of the disk, and {\em r} is the radial distance to the disk center. Applying the Gauss theorem, one can obtain the gravitational potential created by the disk as

\begin{equation}
\centering
\Phi = -\beta \frac{GM_d}{\sqrt{r^2 + r_d^2}}.
\end{equation}

This potential generated by the disk adds one term to the expression of the stellar collision rate $\tau_{coll}^{-1}$:

 \begin{equation}
\centering
  \frac{1}{\tau_{coll}} [yr^{-1}] = 4\sqrt{\pi}\rho_* \sigma_*  r_{coll}^{2} \left(1+\frac{GM}{\sigma_{*}^{2} r_{coll}}+\frac{\beta GM_d}{\sigma_*^2 \sqrt{r_{coll}^2 + r_d^2}}\right)
   \end{equation}

In this case, a collision in the system would occur when r$_{coll}$ = r$_{disk}$, considerably increasing the probability of collision.

\subsubsection{Capture of the incident star by the system star+disk}

When a collision as described in A.2. is produced, the disk may capture the incident star. The condition for a star to be captured by the disk is that the energy of the incident star is lower than the potential energy created by the star+disk system:

\begin{equation}
\centering
\frac{1}{2}m \sigma^2_*  <  \frac{\beta GM_{d} m}{\sqrt{r_{coll}^2 + r_d^2}} + \frac{GMm}{r_{coll}} ,
\end{equation}

This gives a maximum value of r$_{coll}$ for a star to be disk-captured in the collision:

 \begin{equation}
\centering
 r_{max} =  \frac{2G}{\sigma^2_*} \left(\frac{\beta M_d}{\sqrt{2}} + M \right).
   \end{equation}

Then, if r$_{coll}$ $<$ r$_{max}$, a binary could be formed as a consequence of the collision.

\subsubsection{Orbital decay in the binary system induced by accretion}

The subsequent accretion onto a binary system will harden the binary, moving the orbit closer. This produces an orbital decay (Bate et al., \cite{bate02}) and eventually could end in a collision between the two stars. The evolution of the semi-mayor axis of the binary orbit ({\em a}) due to accretion can be described as (Rayburn, \cite{rayburn76})

 \begin{equation}
\centering
 \dot{a}= -3 \hspace{1mm} a\frac{\dot{M}}{M} ,
 \end{equation}

where we considered that the two binaries have the same mass M. Integrating this expression from the moment when both stars have initial masses M$_{i}$ and an initial semi-major orbit a$_i$, to the moment when the stars have accreted mass to form stars with mass M$_{f}$, we obtain the value of the final semi-major axis

 \begin{equation}
\centering
 a_f = a_i \left(\frac{M_i}{M_f}\right )^{3} < a_i .
 \end{equation}

\subsection{The OHC stellar cluster}

 \begin{figure*}
   \centering  
   \includegraphics[angle=0,width=14cm]{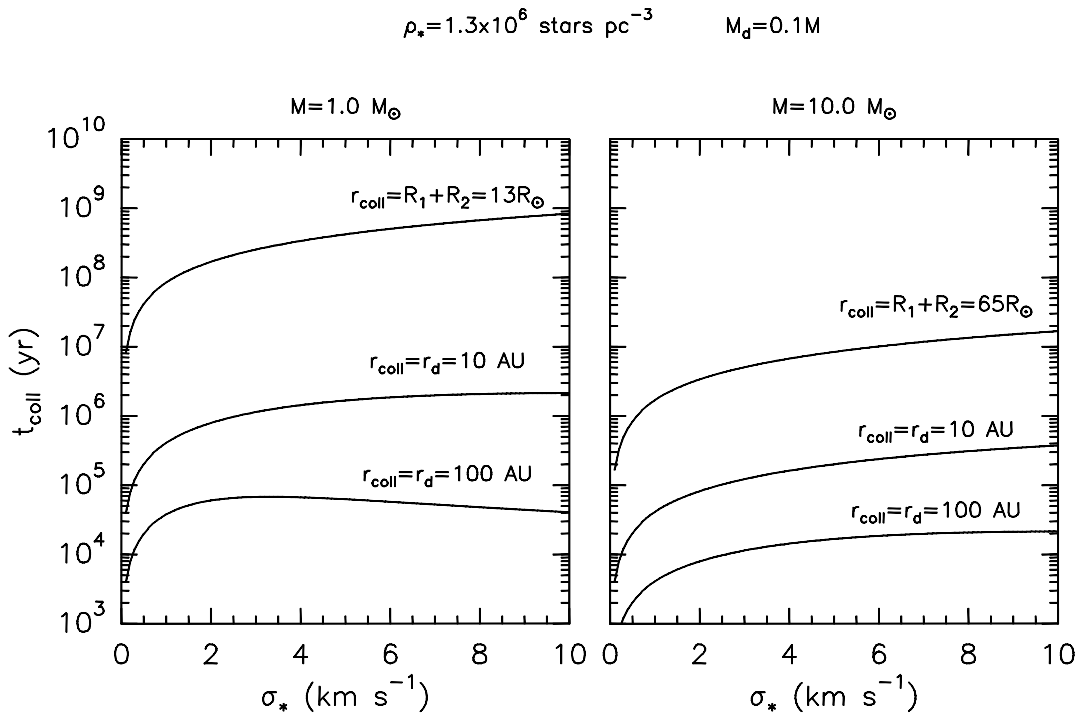}
      \caption{Stellar collision time t$_{coll}$ versus stellar velocity dispersion $\sigma_*$. We have used the lower limit for the stellar density we have determined for the OHC of 1.30$\times$10$^{6}$ stars pc$^{-3}$.  We have considered collisions between 1 M$_{\odot}$ stars (left panel), and 10 M$_{\odot}$ stars (right panel). We have included the case of a direct collision between $^{\prime}$naked$^{\prime}$ stars (eq. A.1.) and the case with the likely presence of circumstellar disks (eq. A.2) with radii 10 and 100 AU (see text).}
         \label{Fig10}
 \end{figure*}

In section A.1. we have derived the expressions for the time needed for a stellar collision to occur in a stellar cluster, t$_{coll}$. Fig. A.1. shows the collision time t$_{coll}$ as a function of the stellar velocity dispersion $\sigma_{*}$ in a cluster of low-mass stars with a stellar density equal to that found in the OHC of 1.30$\times$10$^{6}$ stars pc$^{-3}$. We considered collisions between 1 M$_{\odot}$ stars (left panel), and 10 M$_{\odot}$ stars (right panel). We included the case of a direct collision between $^{\prime}$naked$^{\prime}$ stars (eq. A.1.) and the case with the likely presence of circumstellar disks (eq. A.2.) (Eisner et al., \cite{eisner08}). In the direct $^{\prime}$naked$^{\prime}$ star collision, the collision distance r$_{coll}$ is the sum of the radii of the stars. We considered that the radii of the stars are those of stars with 1 M$_{\odot}$ and 10 M$_{\odot}$ at 10$^{5}$ yr (left and right panels, respectively), the typical timescale for massive star formation. In the case including disks, the collision distance can be as high as the size of the disk (Bally \& Zinnecker, \cite{bally05}). We considered disks with a mass of 0.1 times the mass of the star and reasonable radii of 10 and 100 AU, which can significantly reduce the collision time via disk-captures (Zinnecker \& Yorke, \cite{zinnecker07}; Davies et al., \cite{davies06}; Bonnell $\&$ Bate, \cite{bonnell05}). Moreover, binaries and/or multiple stellar systems are commonly found in Orion (Kraus et al., \cite{kraus09}) and could additionally decrease t$_{coll}$, favoring coalescence.

 There are no direct measurements of the stellar velocity dispersion of the OHC cluster $\sigma_*$. However, lower and upper limits can be estimated. As indicated by Bonnell \& Bate (\cite{bonnell06}), one might expect that the stars forming part of a small-N young cluster like that in the OHC have a low-velocity dispersion. These authors proposed to use values as low as 0.4 km s$^{-1}$. G\'omez et al. (\cite{gomez08}) claimed that typical random motions of recently formed stars have a velocity dispersion of 1-2 km s$^{-1}$. The ONC has a measured velocity dispersion of $\sim$2.3 km s$^{-1}$ (van Altena et al., \cite{vanaltena88}). On the other hand, we can use the gas velocity dispersion of the OHC as a conservative upper limit to the stellar velocity dispersion, because stars are expected to be much less affected by turbulence than gas. The gas velocity dispersion in the OHC, using HC$_{3}$N* (see section 4) is $\sigma$ = $\Delta v$/(2$\sqrt{2\, Ln2}$) $\sim$ 3 km s$^{-1}$ (de Vicente et al., \cite{devicente02}; Rivilla et al., \cite{rivilla13c}, in preparation). This upper limit for $\sigma_*$ is very consistent with the values presented before. 

According to Fig. A.1., a stellar cluster as dense as the OHC with a stellar velocity dispersion of $\sigma_* \sim$ 3 km s$^{-1}$ could experience one stellar collision of two 1 M$_{\odot}$ stars with circumstellar disks with radius 10-100 AU (left panel) in a collision time of about $\sim$10$^5$ yr, making it plausible that the OHC cluster has suffered an encounter of low-mass stars. If a 10 M$_{\odot}$ star (which can be formed by competitive accretion in the low-mass star cluster, see section 4.3) is involved, the collision time is even shorter. Therefore, this analysis suggests that the violent event that has been proposed for the OHC could be a collision between two members of the dense cluster revealed by X-rays studied in our work.

At this point, one may wonder if this close encounter may lead to a direct star-star merger. This is a necessary condition (although it may not be sufficient) to build up a more massive star. Of course, a rigorous characterization of the evolution of this system would need complex simulations that are out of the scope of this appendix. However, we can make some assumptions and simple calculations to determine if this is, at least, viable. 

After one star has impacted the disk of a neighboring star, it can be captured to form a binary if its energy is lower than the potential energy created by the star+disk system (Bally \& Zinnecker, \cite{bally05}; see also section A.1.3.). This condition sets a maximum collision distance at which the star can be captured, r$_{max}$. According to eq. A.6. this maximum distance in a cluster with stellar velocity dispersion of 3 km s$^{-1}$ and disk mass M$_{d}$=0.1M$_{star}$, is $\sim$250 AU for a 10 M$_{\odot}$ star and $\sim$2500 AU for a 1 M$_{\odot}$ star. Therefore, the encounters we considered in this work with r$_{coll}\leq$100 AU can potentially form a binary system as a consequence of the disk capture.

Once the binary system is formed, subsequent accretion of the parental molecular core will harden the binary, triggering an orbital decay (Bate et al., \cite{bate02}). 
Bonnell \& Bate (\cite{bonnell05}) proposed that competitive accretion in a dense cluster naturally forms more massive and closer binaries. If we consider stars with 1 M$_{\odot}$ with disks of reasonable radii of $\sim$100 AU, collisions with r$_{coll}$=r$_{disk}$ are expected to occur in 7$\times$10$^{4}$ yr for a stellar velocity dispersion of $\sigma_*$ $\sim$ 3 km s$^{-1}$ (Fig. A.1.). Fig. 9 shows that the binary system, if located at the center of the OHC low-mass star cluster, can gather a mass of 5 M$_{\odot}$ via competitive accretion in $\sim$2-3$\times$10$^5$ yr. 
Therefore, in several 10$^5$ yr the final individual masses of the binary would be 6 M$_{\odot}$. Then, from eq. A.8. we find that the continuous accretion onto the binary could reduce the semi-major axis of the orbit in a factor 0.005. Considering an initial semi-major axis a$_i$=100-1000 AU, a$_{f}$ would then be 0.5-5 AU. These simple calculations are roughly consistent with the results from numerical simulations of Bonnell \& Bate (\cite{bonnell05}; see their Fig. 2), where the binaries forming in this way typically have very eccentric orbits with periastron separations comparable to the stellar radii, allowing mergers in clusters with stellar densities of 10$^{6}$ stars pc$^{-3}$. The stellar density in the OHC is similar to this value, and the final semi-axis binary separation we estimated can be as low as tens of solar radii, which is slightly higher but on the order of the sum of the radius of two 6 M$_{\odot}$ stars with $\sim$10$^{5}$ yr (Siess et al., \cite{siess00}). This implies that a collision due to orbital decay in a binary system induced by accretion could occur in the OHC.

Whether such a merger leads to a more massive object is difficult to asses. However, we can compare our analysis with other works where stellar coalescence play a role in massive star formation (Davies et al., \cite{davies06} and Moeckel \& Clarke, \cite{moeckel11}). The first work suggests that very high stellar densities (10$^{9}$ M$_{\odot}$ pc$^3$) are needed to form a 50 M$_{\odot}$ star through mergers in the adequate timescale. This is three orders of magnitude higher than what we found in OHC. However, as Davies et al. (\cite{davies06}) recognized in their work, this stellar density can decrease significantly if circumstellar disks or primordial binaries are considered. Indeed, Davies et al. (\cite{davies06}) showed that including circumstellar disks decreases the collision time (and the stellar density needed) by a factor up of 3-10, and that the presence of primordial binaries additionally decreases the stellar density by a factor of 3. In addition, these authors considered the formation of a 50 M$_{\odot}$-star by coalescence, while the most massive objects in the OHC region have masses of only $\sim$10 M$_{\odot}$. According to Fig. 5 in Davies et al. (\cite{davies06}), a star with 10 M$_{\odot}$ is formed in a time $\sim$2.5 shorter than a star with 50 M$_{\odot}$. Then, the stellar density presented in Davies et al. (\cite{davies06}) could be reduced by nearly 2 orders of magnitude, resulting in 10$^{7}$ M$_{\odot}$ pc$^{3}$. This density is still higher than what we found in the OHC, suggesting that coalescence cannot be considered a general process for building up massive stars. This is because to be effective it needs multiple stellar collisions (Davies et al., \cite{davies06}). According to our analysis, a single merger could occur in the cluster evolution, but not many. 

This also agrees very well with the recent work by Moeckel \& Clarke (\cite{moeckel11}). They simulated an Orion-like cluster, assuming a stellar density of 2.5 $\times$10$^4$ stars pc$^{-3}$, and followed its evolution until 3 Myr, the approximate age of the cluster. Even with this relatively low stellar density, in two of their five runs they observed two stellar collisions involving binaries at the central and dense regions of the cluster. 
However, they concluded that an Orion-like cluster can be explained without many collisions. This is evidence against the need for collision to explain massive star formation. Furthermore, even when a collision is involved, Moeckel \& Clarke (\cite{moeckel11}) found that the final mass is mainly determined by accretion growth. Bonnell \& Bate (\cite{bonnell05}) also showed that in collisions the mass accreted by accretion is generally higher than the mass directly gathered by the mergers.

In conclusion, coalescence certainly does not seem a common mechanism for building up massive stars. However, this does not directly rule out the possibility that it may work under certain circumstances, or that a stellar collision may occur in a massive star cradle, especially in regions with high stellar densities. In particular, there is enough evidence to make a stellar collision possible in the OHC. This is supported by the very high stellar density we found in our work ($\sim$10$^{6}$ stars pc$^{-3}$). Whether this collision produced a massive star as a result of a stellar merger still remains unclear, and the current observations do not allow to resolve this controversial question. Moreover, given that other mechanisms such as competitive accretion (see section 4.3.) can explain massive star formation in the OHC, there is no need to invoke coalescence to build up massive stars.

\end{document}